\documentclass[pra,aps,amssymb,twocolumn,noshowpacs,hyperref]{revtex4-1}
\usepackage{color}
\usepackage{graphicx}
\usepackage{hyperref}
\usepackage{amsmath}
\usepackage{latexsym}
\usepackage{amssymb}
\usepackage{mathrsfs}
\usepackage{mathtools}
\usepackage{layout}
\usepackage{verbatim}
\usepackage{multirow}
\usepackage{bm}
\usepackage{amsfonts,epsfig}
\usepackage{lineno}
\usepackage{textcomp}
\usepackage[dvipsnames]{xcolor}
\definecolor{myblue}{RGB}{0, 100, 200}

\begin{document}

\title{Macroscopic position-position entanglement by photon recoil in Rydberg atoms}

\date{\today}
\author{Xiao-Feng Shi}
\affiliation{Center for Theoretical Physics and School of Physics and Optoelectronic Engineering, Hainan University, Haikou 570228, China}
\begin{abstract}
Entanglement between two spatially separate matter particles can be generated via many means and often resides in the internal states of particles. Here, via Rydberg blockade in two spatially separate neutral atoms, we find that the photon recoil in Rydberg excitation can push one atom microns away provided the other atom exerts a state-dependent Rydberg-mediated blockade. When the atoms are recaptured by optical traps, a position-position entangled state between two spatially separate atoms can emerge. This realizes a Bell state of two atoms, where the entanglement exists in the position of each atom and the distance between the two possible locations of each atom can be in the hundred-micron regime.

\end{abstract}
\maketitle

\section{introduction}\label{sec01}

Quantum entanglement in spatially separate particles is considered to be the most nonclassical manifestation of quantum formalism~\cite{Horodecki_2009}. The simplest entangled state between two particles is the Bell state~\cite{Bell_1964}
\begin{eqnarray}
\lvert\Psi_{\text{Bell}}\rangle  &= & \lvert \uparrow_{\text{a}}\downarrow_{\text{b}}\rangle +  \lvert \downarrow_{\text{a}}\uparrow_{\text{b}}\rangle  )/\sqrt{2}.\label{Bell}
\end{eqnarray}
Equation~(\ref{Bell}) can be of several meaning, among which the most studied one  is that a and b are two separate particles and $\uparrow$ and $\downarrow$ are two internal states~\cite{Horodecki_2009,Raimond_2001}. This is because for entanglement to deterministically appear, usually an internal-state dependent interaction is required.

Entanglement can also exist in external states. For example, by using tunnelling of atoms in optical superlattices, an NOON state $\lvert \text{n0}\rangle +  \lvert \text{0n}\rangle  )/\sqrt{2}$ of the positions of an atomic cluster in neighboring lattice sites can be prepared, where the atomic cluster is formed with $n$ indistinguishable atoms from a Bose-Einstein condensate, with $n$ up to seven~\cite{Zhang_2026}. This state is a superposition of the atomic cluster being in two neighboring lattice sites, where the cluster behaves as one particle~\cite{Bloch_2008}. Similarly, an atomic BEC cloud with different internal spin states can favor different regions in an optical trap, rendering entanglement between the cloud's internal states and external residing locations inside the trap~\cite{Lange_2018}, which is still a superposition of states of one material atomic cloud. Entanglement can be prepared between different motional states of two atoms~\cite{Shaw_2025}, where the different motional states are spatially overlapping.

Here, we study another type of deterministic entanglement generation between two spatially separate particles, where $\uparrow_{\alpha}$ and $\downarrow_{\alpha}$ are two different possible positions of particle $\alpha=$a or b and $\uparrow_{\text{a}},\downarrow_{\text{a}}\neq \uparrow_{\text{b}}, \downarrow_{\text{b}}$. More importantly, the spatial distance $\ell$ between $\uparrow_{\alpha}$ and $\downarrow_{\alpha}$ can be in the sub-mm regime. Such a state is like saying that Alice and Bob are entangled in a way that if at a certain moment we find Alice on the north pole of the earth, then Bob must be on the south pole of the moon, but if at that moment we find Alice on the the north pole of the moon, then Bob is on the south pole of the earth. Such a macroscopic position-position entanglement can be generated with photons in phase-matched parametric down conversion~\cite{Howell_2004}, however, in a probabilistic manner.

A deterministic generation of the exotic position-position entanglement of individual atoms can be via photon recoil in Rydberg excitation of neutral atoms. For two atoms labeled c and t, where ``c'' and ``t'' stand for ``control'' and ``target'', when atom c is~(isn't) excited to Rydberg state, the nearby atom t can~(won't) be blocked in a following attempt of Rydberg excitation~\cite{Saffman2010}. Once excited to Rydberg state, atom t will attain a momentum kick $\hbar \mathbf{k}$ with $\mathbf{k}$ the wavevector of the Rydberg laser. A subsequent Rydberg laser with a reverse propagation direction, i.e., with a wavevector $-\mathbf{k}$ can restore atom t back to the ground state, resulting in a net momentum kick $2\hbar \mathbf{k}$ in atom t provided atom c has never been Rydberg excited. For alkali-metal atoms Cs~(Rb), a two-photon Rydberg excitation with copropagating laser fields can yield a $\mathbf{k}$ over 19~(21)$~\mu$m$^{-1}$, so that a flight time up to $100~\mu$s can yield a position change $\ell$ of several microns for the atomic location. If similar laser fields are used for atom c in a superposition of two internal states $\lvert0(1)\rangle$ where only $\lvert1\rangle$ can be Rydberg excited, then a position-position entanglement can emerge. As shown later, a spatial separation $\ell$ around $0.1$~mm is realizable, making macroscopic position-position entanglement possible, which introduces another dimension to the Rydberg-mediated quantum science~\cite{PhysRevLett.85.2208,Lukin2001,Saffman2010,Saffman2016,Saffman2019nsr,Shi2021qst,Browaeys2020,Adams2020,Morgado2021,Wu2021,Kaufman2021,Shao2024review}.

\section{Entanglement between one atom's internal state and another atom's position}\label{sec02}

Before studying the entanglement between the positions of two separate atoms, we show a protocol to generate entanglement between one atom's internal state and another atom's position. 

The entanglement of real-space position of atoms involves the recapture of the atoms after free flight, which is easier with atoms initially cooled to the motional ground state in shallow traps. Atoms are initially trapped with trap frequencies $(\omega_x,~\omega_y,~\omega_z)$ by trap light along $\mathbf{z}$. When the traps are released, the Gaussian width of the atomic wavepacket $\sigma_{j0}=\sqrt{\hbar/(2m\omega_j)}$ will evolve to $\sigma_j = \sigma_{j0}\sqrt{1+ \omega_j^2t^2}$~\cite{Verstraten_2025}, where $j\in\{x,y,z\}$. We consider traps with $\omega_x,~\omega_y$ and $\omega_z$ of similar order of magnitude, and $\omega_z$ on the order of $2\pi\times10~$kHz. In this case, the size change of the Gaussian wavepacket, $\sigma_{j}/\sigma_{j0}$, is negligible with a flight time $\mathbb{T}$ on the order of $10~\mu$s, and the success probability to recapture the atoms by the same optical tweezer light will be high. If $\mathbb{T}$ of order of 100~$\mu$s is needed, stronger tweezer light can be employed to recapture the atoms. Below, we assume that the recapture of the atomic wavepacket has unit efficiency so as to focus on the photon recoil-induced entanglement. In this case, the external state of the atom, i.e., the spatial wavepacket of the atom, can be written as $\lvert \psi_0(\mathbf{r})\rangle$, where $\mathbf{r}$ is the center of the mass of the wavepacket, and the subscript 0 denotes that the expected momentum of the wavepacket is zero. With free-flight times up 100~$\mu$s, the gravitation-induced atomic speed can be neglected.

Consider two internal states $\lvert0\rangle$ and $\lvert1\rangle$. In the experiment of atom-based quantum computing, these two states can be defined in several ways. If alkali-metal atoms are used, they can be two hyperfine-Zeeman substates~\cite{Wilk2010,Isenhower2010,Zhang2010,Maller2015,Jau2015,Zeng2017,Levine2018,Picken2018,Omran2019,Levine2019,Graham2019,Jo2019,Fu2022,McDonnell2022,Bluvstein2022,Graham2022,Evered2023,bluvstein2025,Pan2026atom,evered_high-fidelity_2026,Saffman2026}; if an alkaline-earth like atom is used, they can be fine-structure states~\cite{Unnikrishnan_2024}, optical clock qubits~\cite{Kim_2025,Madjarov2020,tsai2024fid,Finkelstein2024,velocity2026}, or nuclear-spin states~\cite{Ma2022,Ma2023,Muniz2025,Peper2025,senoo2025}. Below, when a ket has two digits, they represent the product state of the internal states of the two atoms c and t.

We consider an initial state $\lvert\Psi_{\text{ct}}\rangle\equiv\lvert\Psi_{\text{c}}\rangle\otimes \lvert\Psi_{\text{t}}\rangle$ of the atoms c and t which are at $\mathbf{r}_{\alpha}=(x_{\alpha0},y_{\alpha0},z_{\alpha0})$ with $\alpha=$ c and t, where
\begin{eqnarray}
\lvert\Psi_{\text{c}}\rangle  &= & \lvert \psi_0(\mathbf{r}_{\text{c}})\rangle\otimes
(\lvert 0\rangle + \lvert 1\rangle  )/\sqrt{2},
\lvert\Psi_{\text{t}}\rangle  =\lvert \psi_0(\mathbf{r}_{\text{t}})\rangle\otimes
\lvert 1\rangle.  \nonumber\label{twoTran}
\end{eqnarray}
The expansion of $\lvert \psi(\mathbf{r})\rangle$ during free flight is assumed though it is not explicitly written. At $t=0$, the trap of atom t is switched off, and a $\pi$ pulse of laser field of Rabi frequency $\Omega$ is sent to atom t which induces a resonant ground-Rydberg transition $\lvert1\rangle\rightarrow \lvert r\rangle$, so that the state of atom c becomes~\cite{Shi2020prapplied}
\begin{eqnarray} \big[\lvert
( \psi_0(\mathbf{r}_{\text{c}})\rangle\otimes\lvert 0\rangle -i e^{ik_{\text{c}}(z_{\text{c}0}+v_{\text{c}}t_\pi/2)} \lvert \psi_{\hbar k_{\text{c}} }(\mathbf{r}_{\text{c}}) \rangle \otimes\lvert r\rangle  \big]/\sqrt{2}
,\label{twoTran2}
\end{eqnarray}
 at $t_\pi=\pi/\Omega$, where $k_{\text{c}}$ is the wavevector of the laser excitation, and $\hbar k_{\text{c}}$ is the momentum kick. The equation above is representative because in the context of wavepacket, the atom is spreading in the three dimensions, while the phase factor $e^{ik_{\text{c}}(z_{\text{c}0}+v_{\text{c}}t_\pi/2)}$ is with one velocity, which should be numerically treated by sampling over the distribution of the velocity~\cite{Shi2020prapplied}. This latter phase factor is the cause of Doppler effect~\cite{Shi2020prapplied}, which is included for completeness, but not essential. This is because the Doppler phase is with the internal state, and there is no need to have a definite phase in the internal state for the desired entanglement to emerge~\footnote{In tweezer-based neutral atom quantum computing, the phase of the qubit state is supposed to be definite, where the Doppler phase is a noise~\cite{Shi2020prapplied}, in contrast to the case here.}. However, the Doppler effect causes error to restore the internal state of the atoms as analyzed later.

 Soon after the laser excitation of atom c, we switch on a laser field which induces the Hamiltonian
\begin{eqnarray}
\hat{H}_{\text{t}}&=& \Omega e^{ik_{\text{t}}(z_{\text{t}0}+v_{\text{t}}t) }\lvert r\rangle\langle 1\rvert/2,~~t\in(t_\pi,~2t_\pi),
\label{twoTran3}
\end{eqnarray}
 for atom t, where $k_{\text{t}}$ is the wavevector of the laser excitation and $v_{\text{t}}$ is representative again. After a $\pi$ pulse of Eq.~(\ref{twoTran3}), another $\pi$ pulse is used, but with opposite laser propagation, i.e., with a Hamiltonian
\begin{eqnarray}
\hat{H}_{\text{t}}&=& \Omega e^{-ik_{\text{t}}(z_{\text{t}0}+v_{\text{t}}t) } \lvert r\rangle\langle 1\rvert/2,~~t\in(2t_\pi,~3t_\pi)
.\label{twoTran4}
\end{eqnarray}
When there is a blockade interaction $V$ in $\lvert rr\rangle$ and when atom c is in the Rydberg state, the two resonant $\pi$ pulses on atom t barely cause any Rydberg excitation~\cite{Saffman2010}, with a residual blockade leakage $\epsilon_{\text{leak}}\leq \hbar^2\Omega^2/(2V^2)$~\cite{Shi2018prapp2}\footnote{It was shown in Ref.~\cite{Shi2018prapp2} that the error is about $\hbar^2\Omega^2/V^2$ times a sine factor, where the sine factor is a function of $V\pi/(\hbar\Omega)$. Given that $V$ can fluctuate, one can see that the the error is bounded below $\hbar^2\Omega^2/V^2$. Half of the final state does not experience the blockade, so the average blockade error is  bounded below $\hbar^2\Omega^2/(2V^2)$.}. But if atom c is at the ground state, Eqs.~(\ref{twoTran3}) and~(\ref{twoTran4}) will excite atom t to Rydberg state and back again, each with a momentum kick $\hbar k_{\text{t}}$. At $t=3t_\pi$, the same laser field for exciting atom c to Rydberg state is used, with a $\pi$ pulse again, which deexcites atom c back to the ground state. Because the laser fields for exciting and deexciting atom c are the same, without any change of propagation direction, there is no net momentum kick in atom c. As a result, the state of the atoms becomes
\begin{eqnarray}
\lvert\Psi_{\text{ct}}(4t_\pi)\rangle  &= &\frac{-1}{\sqrt{2}} [e^{i\varphi_0}\lvert \psi_0(\mathbf{r}_{\text{c}})\rangle\otimes \lvert \psi_{2\hbar k_{\text{t}}}(\mathbf{r}_{\text{t}})\rangle\otimes
\lvert 01\rangle \nonumber\\
& &+  e^{i\varphi_1} \lvert \psi_0(\mathbf{r}_{\text{c}})\rangle\otimes \lvert \psi_0(\mathbf{r}_{\text{t}})\rangle\otimes
\lvert 11\rangle ],\label{twoTran5}
\end{eqnarray}
where $\varphi_0=2k_{\text{t}}z_{\text{t}0}+3k_{\text{t}}v_{\text{t}}t_\pi$ and $\varphi_1=-3k_{\text{c}}v_{\text{c}}t_\pi$~\cite{Shi2020prapplied} arise from Doppler dephasing. In Eq.~(\ref{twoTran5}), the position change of the atomic wavepacket due to the photon recoil is not written for brevity. This change is small if we take, e.g., the two-photon excitation configuration with the $6P_{3/2}$ intermediate state of $^{87}$Rb~\cite{evered_high-fidelity_2026}, then the atom flies by about 0.5~(0.2)~nm with $k_{\text{t(c)}}=2\pi\cdot(1/420\pm1/1015)$nm$^{-1}$ in a flight time $t_\pi\approx33$~ns if we use an experimentally available $\Omega=2\pi\times15$~MHz~\cite{evered_high-fidelity_2026,Muniz2025}.

When atom c is excited back to the ground state, the trap light for it is switched on, while that for atom t is not, and, hence, atom t flies freely. For the case of the Rydberg excitation via the $6P_{3/2}$ intermediate state of rubidium as above, we have $\mathbb{V}=2\hbar k_{\text{t}}/m\approx30.9$~nm$/\mu$s. Because all the internal states of the two atoms are in the ground state, and the coherence time of the internal qubit states in atom c can be on the order of 10~s~\cite{6100atoms} or even over 100~s~\cite{Kim_2025}, the free flight of atom t with a time $100~\mu$s barely decoheres the internal state. The wavepacket of atom t with a momentum $2\hbar k_{\text{t}}$ will induce a position change of its center of mass by $\ell\approx3.1~\mu$m after $\mathbb{T}=100~\mu$s, and, then, two trap light beams can be switched on for atom t, one focusing on $\mathbf{r}_{\text{t}}+ \ell\mathbf{e}_{\text{z}}$, and the other focusing on $\mathbf{r}_{\text{t}}$. Once the atoms are trapped, the subscript of $\psi$ denoting the momentum of the atoms can be dropped, so that
\begin{eqnarray}
\lvert\Psi_{\text{ct}}(4t_\pi+\mathbb{T})\rangle  &= &\frac{-1}{\sqrt{2}} [e^{i\varphi_0}\lvert \psi(\mathbf{r}_{\text{c}})\rangle\otimes \lvert \psi(\mathbf{r}_{\text{t}}+ \ell\mathbf{e}_{\text{z}})\rangle\otimes
\lvert 01\rangle \nonumber\\
& &+  e^{i\varphi_1} \lvert \psi(\mathbf{r}_{\text{c}})\rangle\otimes \lvert \psi(\mathbf{r}_{\text{t}})\rangle\otimes
\lvert 11\rangle ],\label{twoTran6}
\end{eqnarray}
which is entangled in that if the internal state of atom c is $\lvert0(1)\rangle$, then atom t is at $\mathbf{r}_{\text{t}}+ \ell\mathbf{e}_{\text{z}}$~($\mathbf{r}_{\text{t}}$). The phase $\varphi_0$ is random, but the lifetime of the entangled state in Eq.~(\ref{twoTran6}) is limited by the relaxation time $T_1$ of the internal state $\lvert0(1)\rangle$ of atom c~\footnote{The quoted times from Refs.~\cite{Kim_2025,6100atoms} are the coherence times. However, the relaxation time $T_1$ is longer than the coherence time of the qubit for the atoms under study because the coherence between the two qubit states includes the relaxation effect. }, which can be over 10~s~\cite{6100atoms} or even 100~s~\cite{Kim_2025}. Note that though the position state of the atom is long-lived, with a lifetime well over 1000~s~\cite{6100atoms,Schymik_2021,Zhang_2025} limited by the residual vacuum pressure, the relaxation of the internal state is observable by the environment. In other words, when the relaxation $\lvert0\rangle\leftrightarrow\lvert1\rangle$ occurs, the environment has detected the atomic positions, instantly erasing the entanglement.

\section{Position entanglement between two separate atoms}\label{sec03}

A slight variation of the protocol above can be used to create entanglement between the positions of two atoms. To prepare Eq.~(\ref{twoTran6}), the laser field for atom c differs from that for atom t. But if we use a similar configuration of field propagation for both atoms, then the first and fourth laser pulses are like those for Eqs.~(\ref{twoTran3}) and~(\ref{twoTran4}), i.e., the Hamiltonian for atom c is
\begin{eqnarray}
\hat{H}_{\text{c}}&=& \left\{\begin{array} {ll}\Omega e^{\pm ik_{\text{c}}(z_{\text{c}0}+v_{\text{c}}t) }\lvert r\rangle\langle 1\rvert/2,&~~t\in(0,~t_\pi),\nonumber\\
\Omega e^{\mp ik_{\text{c}}(z_{\text{c}0}+v_{\text{c}}t) }\lvert r\rangle\langle 1\rvert/2,&~~t\in(3t_\pi,~4t_\pi),
                                                        \end{array}\right.
\label{se03e01}
\end{eqnarray}
where the signs $\pm$ and $\mp$ denote that as long as the directions of the laser fields are reversed, there will be position-position entanglement between the two atoms finally.

Beside of the change of the laser field to atom c, a second change is that after the fourth laser pulse, atom c is also left to freely fly for the same time as atom t. Consider a similar two-photon Rydberg excitation as below Eq.~\ref{twoTran5}, i.e., let $k_{\text{c}}=k_{\text{t}}\approx 21\mu$m$^{-1}$, the change of $z_{\text{c}}$ would be the same to that of atom t. As a result, when the tweezer light for trapping atom t is switched on at $t=4t_\pi+\mathbb{T}$, similar tweezer light should be switched on for atom c, focusing on $\mathbf{r}_{\text{c}}\pm \ell\mathbf{e}_{\text{z}}$ and $\mathbf{r}_{\text{c}}$. Then, the state of the two atoms becomes
\begin{eqnarray}
\lvert\Psi_{\text{ct}}(4t_\pi+\mathbb{T})\rangle  &= &\frac{-1}{\sqrt{2}} [e^{i\varphi_0}\lvert \psi(\mathbf{r}_{\text{c}})\rangle\otimes \lvert \psi(\mathbf{r}_{\text{t}}+ \ell\mathbf{e}_{\text{z}})\rangle\otimes
\lvert 01\rangle \nonumber\\
& &+  e^{\pm i\varphi_1'} \lvert \psi(\mathbf{r}_{\text{c}} \pm \ell\mathbf{e}_{\text{z}})\rangle\otimes \lvert \psi(\mathbf{r}_{\text{t}})\rangle\otimes
\lvert 11\rangle ],\nonumber
\label{se03e02}
\end{eqnarray}
where $\varphi_1'=2k_{\text{c}}z_{\text{c}0}+4k_{\text{c}}v_{\text{c}}t_\pi$~\cite{Shi2020prapplied}.

\section{ Faster entanglement with multiple Rabi cycles--}
If the position-position entanglement of two atoms is used in a hybrid quantum system like, e.g., cavity QED system for quantum networking, then the time efficiency to generate the entanglement is important. The example above is not efficient for it needs $100~\mu$s to create entanglement between spatially separate atoms with each atom ``split apart'' by about 3$~\mu$m. This mainly originates from the small drift speed $\mathbb{V}$. There are two ways to enhance it, by more than one Rabi cycles , or by using lighter atoms.

We first analyze using more than one Rabi cycle. Consider using $\mathbb{N}$ Rabi cycles for each atom, then the Hamiltonians for the two atoms are
\begin{widetext}
\begin{eqnarray}
\hat{H}_{\text{c}}&=& \left\{\begin{array} {ll}\Omega e^{\pm ik_{\text{c}}[z_{\text{c}0}+v_{\text{c}}t+ \delta(t)] }\lvert r\rangle\langle 1\rvert/2,&~~t\in(0,~t_\pi),~\text{and~~}t\in(2\mathbb{N}+2j+2,~2\mathbb{N}+2j+3)t_\pi,\nonumber\\
\Omega e^{\mp ik_{\text{c}}[z_{\text{c}0}+v_{\text{c}}t+ \delta(t)] }\lvert r\rangle\langle 1\rvert/2,&~~t\in(2\mathbb{N}+2j+1,~2\mathbb{N}+2j+2)t_\pi,
                                                        \end{array}\right.\nonumber\\
\hat{H}_{\text{t}}&=& \left\{\begin{array} {ll}\Omega e^{ ik_{\text{t}}[z_{\text{t}0}+v_{\text{t}}t+ \delta(t)] }\lvert r\rangle\langle 1\rvert/2,&~~t\in(2j+1,~2j+2)t_\pi,\nonumber\\
\Omega e^{- ik_{\text{t}}[z_{\text{t}0}+v_{\text{t}}t+ \delta(t)] }\lvert r\rangle\langle 1\rvert/2,&~~t\in(2j+2,~2j+3)t_\pi,
                                                        \end{array}\right.
\label{se04e01}
\end{eqnarray}
\end{widetext}
where $j=0,1,2~\cdots,N-1$ and $\delta(t)$ is the change of the position along $\mathbf{z}$ due to the velocity gain from the momentum kick. If similar laser excitation scheme is used for both atoms, then $k_{\text{c}}=k_{\text{t}}\equiv k$. Each pair of Rydberg excitation and deexcitation will incur a momentum kick $2\hbar k$, so that for $\mathbb{N}$ Rabi cycles, the total momentum kick is enhanced by $\mathbb{N}$ times compared to using only one Rabi cycle. If we still use the rubidium example analyzed around Eq.~(\ref{twoTran5}), the drift speed $\mathbb{V}$ would be about 0.31$\mu$m$/\mu$s if $\mathbb{N}=10$, and a flight time $\mathbb{T}= 10~\mu$s would yield a spatial splitting $\ell\approx3.1~\mu$m for each atom.

\begin{figure}
\includegraphics[width=3.4in]
{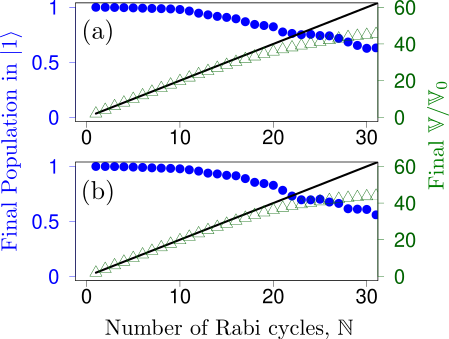}
\caption{Influence of Doppler effect on the restoration of the internal states before free flight by using multiple Rabi cycles with $^{87}$Rb. The round and triangular symbols in (a) and (b) show the final population in the internal state $\lvert1\rangle$ and the final kick-induced atomic velocity $\mathbb{V}$, where $\mathbb{V}_0$ is the velocity gain of the atom in one transition of $\lvert0\rangle\rightarrow\lvert1\rangle$, and (a) and (b) are for the target and control atoms, respectively. The tilted line shows the desired total number of momentum kick, which should be $2\mathbb{N}$ since there are two momentum kicks in one full Rabi oscillation. Here, the data in (b) is with ten Rabi cycles for exciting the target atom, and perfect blockade and absence of Rydberg-state decay are assumed; see text for analysis on the fidelity with blockade and Rydberg-state decay.  \label{figure01} }
\end{figure}

The weakness of $\mathbb{N}\geq2$ is that Rydberg-state decay and the Doppler effect would hamper the population transfer between the internal states $\lvert0\rangle$ and $\lvert1\rangle$~\cite{Shi2020prapplied}, which would be more and more detrimental with larger $\mathbb{N}$~\cite{Shi_2025}. The error due to Rydberg-state decay is $\epsilon_{\text{decay}}= 2\mathbb{N}t_\pi/\tau$ averaged over the two atoms, where $\tau$ is the Rydberg-state lifetime.

There are three sources for Doppler dephasing. First, the random $v$ can be nonzero, leading to the the usual Doppler effect~\cite{Shi2020prapplied}. A proper treatment is to sample $v$ with a Gaussian distribution of width $v_{\text{rms}}$, where $v_{\text{rms}}$ satisfies $mv_{\text{rms}}^2/2=E/2$ with $E=\hbar\omega_z/2$ the energy of the motional ground state and the factor $1/2$ comes from that the average potential energy and kinetic energy are equal in the trap. Second, the random $z_{\text{c(t)0}}$ will cause a Doppler-like dephasing. We can sample $z_{\text{c(t)0}}$ with a Gaussian distribution of width $\sigma_{z0}$ to include its effect. This dephasing is absent in usual Rydberg quantum science where the excitation and deexcitation of the atom is by the same laser fields, so that the phase term from the random $z_{\text{c(t)0}}$ is cancelled. Third, the term $\delta(t)$ also leads to change of the position of atom continuously, causing an extra Doppler shift. If we assume that the absorption or emission of a photon in the Rabi cycles is not instantaneous, then in a numeric study, $\delta(t)$ is given by $\int_0^t \mathbb{V}(\tau)d\tau$, which can be calculated step by step in that the change of the atomic speed is proportional to both $k$ and the population change in the internal state of the atom.

By including the three Doppler processes above, we have numerically studied the quality of the position-position entanglement between the atoms, shown in Fig.~\ref{figure01}. Figure~\ref{figure01}(a) shows that the final population in $\lvert1\rangle$ in the target atom is larger than 0.99 if $\mathbb{N}\leq6$.  Here, we assume a copropagating two-photon excitation configuration with the $6P_{3/2}$ intermediate state of rubidium $k=k_{\text{t(c)}}=2\pi\cdot(1/420+1/1015)$nm$^{-1}$ and $\Omega=2\pi\times15$~MHz~\cite{evered_high-fidelity_2026}, and shallow traps with $\omega_z=2\pi\times10$~kHz. The data are sampled by taking Gaussian distribution of the initial values of $v_\alpha$ and $z_{\alpha0}$. The data in Fig.~\ref{figure01}(a) and (b) differ from each other because in Fig.~\ref{figure01}(b), after exciting atom c to Rydberg state, there is an idling time left for it to fly during which atom t is excited by laser fields. After the wait time, there comes the remaining $2\mathbb{N}-1$ $\pi$ pulses for atom c.

The fidelity to prepare the spatial entanglement can be defined by $\mathcal{F}= p_{\text{c}}p_{\text{t}}(1-\epsilon_{\text{decay}})(1-\epsilon_{\text{leak}})$, where $p_{\text{c(t)}}$ is the population restored after the Rabi cycles for atom c(t). The data in Fig.~\ref{figure01} gives $p_{\text{c(t)}}=0.9801~(0.9782)$ when $\mathbb{N}=10$ for both atoms. If we assume $\tau=80~\mu$s and $V/\hbar=2\pi\times0.95$~GHz as in the experiment of Ref.~\cite{evered_high-fidelity_2026}, then $\mathcal{F}=0.9507$. Compared to $\mathcal{F}=0.9990$ when $\mathbb{N}=1$ for both atoms, using $\mathbb{N}=10$ enhances the time efficiency with fidelity dropped by 5\%.

\section{ Faster entanglement with lighter atoms}
One can use lighter neutral atoms for enhancing the time efficiency, like lithium, sodium, magnesium, and potassium. Among these, the lightest is lithium, which has been experimentally prepared in optical tweezer array~\cite{Spar_2022} and Rydberg excited~\cite{Saakyan_2023}. However, even lighter atoms can be used. For example, Ref.~\cite{JPCovey2026} gave detailed analyses on using the metastable helium-3~($^3$He$^\ast$)~\footnote{Here, $^3$He$^\ast$ instead of $^3$He is used to distinguish that the atom we take here is not in the ground state but in the $2~^3S_1$ state.} for Rydberg atom quantum science, suggesting a two-photon transition,~\cite{JPCovey2026} $1s2s~^3S_1\xrightarrow{389~\text{nm}}
1s3p~^3P_J\xrightarrow{785~\text{nm}}
1sns~^3S_1$. The mass about 3.016~a.u. of $^3$He$^\ast$ is much smaller than the 86.909~a.u. of $^{87}$Rb, and the velocity gain of the $^3$He$^\ast$ atom after one Rabi cycle of Rydberg excitation and deexcitation by opposing copropagating lasers is $\mathbb{V}\approx1.02$~$\mu$m$/\mu$s, so that a flight time $\mathbb{T}=100~\mu$s can yield $\ell\approx 102~\mu$m, making it feasible to generate macroscopic position-position entanglement between two separate neutral atoms with a high efficiency.

\section{Discussion}
Beside of being useful for fundamental physics, the feasibility to generate entanglement between two spatially separate particles' external positions can provide alternative means for quantum networking. For example, by trapping atoms in cavities, the position entanglement of atoms can be transferred to photonic qubits in optical fibres, resulting in entanglement in photons' positions. Consider four fibres $c_{1(2)}$ and $t_{1(2)}$ which have received the entanglement, where $c_{1(2)}$ and $t_{1(2)}$ are for photons from atom c and t, respectively. Then if fibre $c_{1(2)}$ has a photon, then fibre $t_{2(1)}$ has a photon. Importantly, this entanglement is deterministic, in contrast to stochastic position-position entanglement of photons generated in spontaneous emission~\cite{Howell_2004}. This deterministic photon-photon entanglement with qubits defined by existing or absence of photons render an alternative route towards quantum networking.

\section{Conclusions}\label{sec07}
We show that by using photon recoil in Rydberg atoms, a position-position entanglement between two spatially separate ground-state atoms can be realized. The entanglement resides in the real-space positions of the atoms. With frequently used heavy atoms like rubidium, a flight time of order $100~\mu$s is needed to realize micron-scale entanglement size if only one Rabi cycle is used. If more Rabi cycles are used, the time efficiency can be greatly increased. Alternatively, one can use lighter atoms like $^3$He$^\ast$ to enhance the time efficiency for generating the macroscopic spatially entangled states. The introduced position-position entanglement of spatially separate atoms here can be useful for quantum technologies involving atoms.

\section*{acknowledgments}
The author acknowledges the National Natural Science Foundation of China under Grant Nos. 12074300 and 12547103, and the Quantum Science and Technology-National Science and Technology Major project Grant No. 2021ZD0302100, and thanks Yan Lu for discussions.

%

\end{document}